\documentstyle[12pt,equation,epsfig]{article}
\begin{document}
\newcommand{\be}{\begin{equation}}
\newcommand{\ben}{\begin{subequations}}
\newcommand{\een}{\end{subequations}}
\newcommand{\beq}{\begin{eqalignno}}
\newcommand{\eeq}{\end{eqalignno}}
\newcommand{\ee}{\end{equation}}
\newcommand{\mumu}{\mbox{$\mu^+ \mu^-$}}
\newcommand{\tanb}{\mbox{$\tan \! \beta$}}
\newcommand{\mhpl}{\mbox{$m_{H^+}$}}
\newcommand{\stau}{\mbox{$\widetilde \tau$}}
\newcommand{\tchi}{\mbox{$\tilde \chi$}}
\newcommand{\ttau}{\mbox{$\tilde \tau$}}
\newcommand{\rs}{\mbox{$\sqrt{s}$}}
\renewcommand{\thefootnote}{\fnsymbol{footnote}}

\pagestyle{empty}
\begin{flushright}
IFT--P.056/98 \\
YUMS 98--13\\
August 1998\\
\end{flushright}
\vspace*{2cm}
\begin{center}
{\Large \bf Signals for CP--Violation in Scalar Tau Pair Production at
Muon Colliders} \\
\vspace*{6mm}
Seong--Youl Choi$^1$ and Manuel Drees$^2$ \\
$^1${\it Dept. of Physics, Yonsei Univ., Seoul 120--749, Korea} \\
$^2${\it IFT, Univ. Estadual Paulista, 900--01405 S\~ao Paulo, Brazil}
\end{center}

\vspace*{1cm}
\begin{abstract}
We discuss signals for CP--violation in $\mumu \rightarrow \ttau_i^-
\ttau_j^+$, where $i,j = 1,2$ label the two scalar $\tau$ mass eigenstates.
We assume that these reactions can proceed through the production and
decay of the heavy neutral Higgs bosons present in supersymmetric models.
CP--violation in the Higgs sector can be probed through a rate asymmetry
even with unpolarized beams, while the CP--odd phase associated with the
\ttau\ mass matrix can be probed only if the polarization of at least one
beam can be varied. These asymmetries might be ${\cal O}(1)$.

\end{abstract}

\vspace*{3mm}

\clearpage
\pagestyle{plain}
\setcounter{page}{1}

In the past few years a considerable amount of effort has been devoted
to investigations of the physics potential of high energy \mumu\
colliders (MC) \cite{1}. Since muons emit far less synchrotron
radiation than electrons do, an MC might be significantly smaller and
cheaper than an $e^+ e^-$ collider operating at the same
center--of--mass energy \rs. The main physics advantage of MC's is
that the larger Yukawa coupling of muons in many cases admits copious
production of Higgs bosons as $s-$channel resonances, allowing to
perform precision measurements of their properties \cite{1,2,3}. In
particular, one can search for CP--violation in the couplings of Higgs
bosons to heavy Standard Model (SM) fermions \cite{4}.

In this Letter we point out the possibility of studying CP--violating
phases associated with soft supersymmetry breaking at an MC.
Supersymmetry is now widely regarded to be the most plausible
extension of the SM; among other things, it stabilizes the gauge
hierarchy \cite{6} and allows for the Grand Unification of all known
gauge interactions \cite{7}. Of course, supersymmetry must be (softly)
broken to be phenomenologically viable. In general this introduces
a large number of unknown parameters, many of which can be complex.
CP--violating phases associated with sfermions of the first and, to a
lesser extent, second generation are severely constrained by bounds on
the electric dipole moments of the electron, neutron and
muon. However, it has recently been realized \cite{9} that
cancellations between different diagrams allow some combinations of
these phases to be quite large. Even in models with universal boundary
conditions for soft breaking masses at some very high energy scale,
the relative phase between the supersymmetric higgsino mass parameter
$\mu$ and the universal trilinear soft breaking parameter $A_0$ can be
${\cal O}(1)$ \cite{10}.  If universality is not assumed, the phases
of third generation trilinear soft breaking parameters are essentially
unconstrained. In fact, there is reason to believe that some of these
phases might be large \cite{worah}, since many proposed explanations
of the baryon asymmetry of the Universe require non--SM sources of CP
violation.

Unfortunately it is difficult to probe these phases through processes
controlled by gauge interactions, where large CP--odd asymmetries can
emerge only if some sfermion mass eigenstates are closely degenerate,
with mass splitting on the order of the decay width, in which case
``flavor oscillations'' can occur \cite{11a,11b,11c}. On the other
hand, even in the minimal supersymmetric extension of the SM, the
MSSM, CP--violating phases can appear at tree level in the couplings
of a single sfermion species to neutral Higgs bosons. These phases can
give rise to large CP--odd asymmetries regardless of sfermion mass
splittings. Here we focus on \ttau\ pair production. Unlike sfermions
of the first two generations, \ttau's generally have sizable couplings
to heavy Higgs bosons even if the latter are much heavier than
$M_Z$. Furthermore, unlike for $\tilde{b}$ and $\tilde {t}$ production
the charge of a produced \ttau\ is usually readily measurable; this
is necessary for the construction of most CP--odd asymmetries.
Finally, in most models sleptons are significantly lighter than
squarks, making it easier to study them at lepton colliders.

Recently it has been realized \cite{12} that CP--violation in the
sfermion sector will lead to mixing between the CP--even ($h, H$) and
CP--odd ($A$) Higgs bosons of the MSSM. Although this is a radiative
effect, it can change certain asymmetries dramatically. Most of this
effect is expected to come from loops involving $\tilde t$ or $\tilde
b$ squarks. Rather than specifying the numerous free parameters of
these sectors, we simply choose a value for the CP--violating $H-A$
mixing mass term $\delta m^2_{H,A}$, within the range found in
ref.\cite{12}. For realistic \ttau\ masses the exchange of the
lightest Higgs boson contributes negligibly to the matrix element, so
that $h-A$ mixing is of little importance for us.

In general the matrix element ${\cal M}$ for $\mumu \rightarrow
\ttau_i^- \ttau^+_j$ receives contributions from $\gamma$ and $Z$ 
exchange as well as from the exchange of the neutral Higgs bosons of
the MSSM. The square of this matrix element for general (longitudinal
or transverse) beam polarization can be computed either using standard
trace techniques (employing general spin projection operators), or
from the helicity amplitudes by a suitable rotation \cite{13} from the
helicity basis to a general spin basis. Both calculations give the same
result:
\beq \label{e1}
\left| {\cal M} \right|^2 &= \frac {e^4 |{\vec k}|^2 \sin^2 \theta} {2 s}
\left[ \left( |V_{ij}|^2 + |A_{ij}|^2 \right) \left( 1 - P_L \bar{P}_L
\right) + 2 \Re e(V_{ij}A^*_{ij}) \left( P_L - \bar{P}_L \right)
\right. \nonumber \\ & \left. \hspace*{2.4cm}
- \left( |V_{ij}|^2 - |A_{ij}|^2 \right) P_T \bar{P}_T \cos( \alpha +
\bar{\alpha}) \right] \nonumber \\
&+ \frac {h_\mu^2} {4 s} \left[ \left( |P_{ij}|^2 + |S_{ij}|^2 \right)
\left( 1 + P_L \bar{P}_L \right) + \left( |S_{ij}|^2 - |P_{ij}|^2 \right)
P_T \bar{P}_T \cos(\alpha - \bar{\alpha})
\right. \nonumber \\ & \left. \hspace*{1.cm}
+ 2 \Re e(P_{ij} S^*_{ij}) \left( P_L + \bar{P}_L \right) -
2 \Im m(P_{ij} S^*_{ij}) P_T \bar{P}_T \sin(\alpha - \bar{\alpha}) \right]
 \\
&+ \frac{ e^2 h_\mu |\vec{k}| \sin \! \theta}{\sqrt{2} s} \left[
\Re e(S^*_{ij} V_{ij}) \left( \bar{P}_L P_T \cos \! \alpha - P_L \bar{P}_T
\cos \! \bar{\alpha} \right)
+ \Im m(S^*_{ij} V_{ij}) \left( P_T \sin \! \alpha + \bar{P}_T \sin \!
\bar{\alpha} \right) 
\right. \nonumber \\ & \left. \hspace*{2.4cm}
- \Re e(S^*_{ij} A_{ij}) \left( P_T \cos \! \alpha + \bar{P}_T \cos \!
\bar{\alpha} \right)
+ \Im m(S^*_{ij} A_{ij}) \left( P_L \bar{P}_T \sin \! \bar{\alpha}
- \bar{P}_L P_T \sin \! \alpha \right)
\right. \nonumber \\ & \left. \hspace*{2.4cm}
+ \Re e(P^*_{ij} V_{ij}) \left( P_T \cos \! \alpha - \bar{P}_T \cos \!
\bar{\alpha} \right)
+ \Im m(P^*_{ij} V_{ij}) \left( P_T \bar{P}_L \sin \! \alpha + \bar{P}_T
P_L \sin \! \bar{\alpha} \right)
\right. \nonumber \\ & \left. \hspace*{2.4cm}
- \Re e(P^*_{ij} A_{ij}) \left( P_L \bar{P}_T \cos \! \bar{\alpha} +
\bar{P}_L P_T \cos \! \alpha \right) 
+ \Im m(P^*_{ij} A_{ij}) \left( \bar{P}_T \sin \! \bar{\alpha} - P_T
\sin \! \alpha \right) \right]. 
\nonumber
\eeq
Here, $\vec{k}$ is the $\ttau^-$ 3--momentum in the center--of--mass
frame, $\theta$ is the scattering angle, $e$ is the QED gauge
coupling, and $h_\mu = g m_\mu / (2 M_W \cos \! \beta)$ determines the
strength of the $\mu$ Yukawa couplings, \tanb\ being the usual ratio
of Higgs vacuum expectation values. $V_{ij}, A_{ij}, S_{ij}$ and
$P_{ij}$ are combinations of coupling factors and propagators,
corresponding to vector, axial vector, scalar and pseudoscalar
couplings to \mumu, respectively. $V_{ij}$ and $A_{ij}$ are
dimensionless and describe $\gamma$ and $Z$ exchange (only the latter
contributes to $A_{ij}$), while $S_{ij}$ and $P_{ij}$ describe Higgs
exchange contributions and have dimension of mass. Explicit
expressions for these quantities will be given elsewhere
\cite{14}. Finally, $P_L$ and $\bar{P}_L$ are the longitudinal
polarizations of the $\mu^-$ and $\mu^+$ beams, while $P_T$ and
$\bar{P}_T$ are the degrees of transverse beam polarization, with
$\alpha$ and $\bar{\alpha}$ being the azimuthal angles between these
polarization vectors and $\vec{k}$. Note that $P_L^2 + P_T^2 \leq 1$
and $\bar{P}_L^2 + \bar{P}_T^2 \leq 1$.

In this notation a CP--transformation corresponds to the simultaneous
exchanges $P_L \leftrightarrow - \bar{P}_L, \ P_T \leftrightarrow
\bar{P}_T$ and $\alpha \leftrightarrow \bar{\alpha}$. Out of the 15
terms appearing in eq.(\ref{e1}), the first five as well as terms 8
through 11 are CP--even, while the remaining 6 terms are CP--odd. Let
$C_n(i,j)$ be the coefficients of these 15 terms (bilinears in
$V_{ij}, A_{ij}, S_{ij}$ and $P_{ij}$ and their complex conjugates).
For the coefficients multiplying CP--even factors (the first group),
only the antisymmetric combinations $[C_n] \equiv C_n(1,2) - C_n(2,1)$
lead to CP--violation through rate asymmetries. In contrast, all
symmetric combinations $\{C_n\} \equiv (C_n(i,j) + C_n(j,i))/2$ of
the coefficients of the second group of terms contribute to CP--odd
polarization or azimuthal angle asymmetries; these can be probed for
three different CP--even final states ($\ttau_i^- \ttau_i^+, \ i=1,2$
and the sum of $\ttau_1^- \ttau_2^+$ and $\ttau_1^+ \ttau_2^-$
production).

We emphasize that CP--odd combinations of all 15 coefficients appearing
in eq.(\ref{e1}) can be extracted independently, {\em if} the polarization
of both beams can be controlled completely. To mention only two examples,
$\{C_6\}$ can be extracted by measuring the difference of cross sections
for $P_L = \bar{P}_L = +1$ and $P_L = \bar{P}_L = -1$; recall that
$|P_L|=1$ implies $P_T=0$. $[C_{14}]$ can be determined by measuring
$\int d \alpha d \bar{\alpha} |{\cal M}|^2 (\cos \! \alpha + \cos \!
\bar{\alpha})$ for $P_L = \bar{P}_T = 1$, adding the same quantity for
$P_T = - \bar{P}_L = 1$ (for a CP--even polarization state), and
anti--symmetrizing in the \ttau\ indices. In this fashion one can
define 9 rate asymmetries $A_R$ and 6 polarization/angle asymmetries
$A_P$; recall that the latter can be studied for three different final
states, leading to a total of 27 different asymmetries!

How many of these asymmetries can actually be measured in practice
depends on the beam energy (which determines how many different final
states $\ttau_i^- \ttau^+_j$ are accessible) and, crucially, on the
extent to which the beam polarization can be controlled. If this is
not possible at all, only the total rate asymmetry $\propto [ C_1 +
C_4]$ can be measured. If the longitudinal polarization can be tuned
but $P_T = \bar{P}_T = 0$, one can in addition determine a rate
asymmetry $\propto [C_2]$ (which however is expected to vanish, since
$C_2$ only involves gauge interactions) and a polarization asymmetry
$\propto \{C_6\}$. All other asymmetries are only accessible if at
least one beam is transversely polarized. Note that asymmetries that
require only one transversely polarized beam can only be measured if
the azimuthal angle of the \ttau's can be reconstructed; this should
be possible fairly efficiently at least on a statistical basis, unless
one is very close to the threshold (in which case the cross section is
quite small anyway).  Asymmetries that are accessible only if $P_T$
and $\bar{P}_T$ are both nonzero only depend on the difference $\alpha
- \bar{\alpha}$, which is independent of $\vec{k}$; this includes the
polarization/angle asymmetry $\propto \{C_7\}$, which is analogous to
the ``production asymmetry'' introduced in ref.\cite{4}.

Some amount of longitudinal
polarization will likely be present automatically, if the muons are
produced from the weak decay of light mesons. This by itself is not
sufficient to measure polarization asymmetries; one has to be able to
tune the beam polarization, which might entail a significant reduction
of the luminosity \cite{3}. Producing transversely polarized beams
will not be easy. Conventional spin rotators used for electron beams
will not be effective, since the magnetic dipole moments of leptons
scale as the inverse of their mass. It might nevertheless be useful
to investigate what additional information might become accessible
with transversely polarized beams.

To that end we present numerical results for a ``typical'' set of MSSM
parameters: $m_A = |\mu| = |A_\tau| = 500$ GeV, gaugino mass $M_2=300$
GeV, $m_{\tilde \tau_L} = 230$ GeV, $m_{\tilde \tau_R}=180$ GeV and
$\tanb=10$. We set all phases to zero, except for that of $A_\tau$
which we take to be 1. The choice of $M_2$ affects our results only
through the Higgs decay widths, which can get significant
contributions from decays into neutralinos and charginos. The ratio of
heavy Higgs boson masses, controlled by $m_A$, and soft breaking
\ttau\ masses has been chosen such that all combinations $\ttau_i^-
\ttau_j^+$ can be produced in the decay of on--shell Higgs bosons.

In Fig.~1 we show results for the total cross sections for \ttau\ pair
production. In this figure we have set $H-A$ mixing to zero, but
introducing a nonzero $\delta m^2_{H,A}$ in the range found in
ref.\cite{12} has little influence on these results. The nontrivial
phase between $\mu$ and $A_\tau$ leads to CP--violation in the
Higgs--\ttau--\ttau\ couplings, so that the exchange of both heavy
Higgs bosons contributes to all three channels. (If CP is conserved,
$A-$exchange only contributes to $\ttau_1 \ttau_2$ production.) However,
the Higgs decay widths ($\sim 1.2$ GeV for both $A$ and $H$) are
significantly larger than the $H-A$ mass difference of 400 MeV, so that
only a single resonance structure is visible in the line shapes. Higgs
exchange contributions completely dominate $\ttau_1 \ttau_2$ production
for $|\rs-m_A| \leq 3$ GeV, while they are at best comparable to
the gauge contributions for $\ttau_1^- \ttau_1^+$  production. Recall
that these results are for the moderately large value $\tanb=10$.
Increasing \tanb\ even further has little effect on the cross sections
near $\rs = m_A$, since the couplings of the heavy Higgs bosons to
\ttau\ and $\mu$ pairs grow $\propto \tanb$ while the total Higgs
decay widths are $\propto \tan^2 \beta$, but broadens the
poles by a factor $\propto \tan^2 \beta$. Moreover, the Higgs
exchange contributions to the matrix element scale essentially
linearly in $|A_\tau|$ as long as $m_A^2 \gg M_Z^2$ and $\tan^2 \beta
\gg |A_\tau/\mu|$. 

In Figs.~2 and 3 we show some ``effective asymmetries'', defined as
products of an asymmetry and the square--root of the relevant cross
section; these determine the integrated luminosity times
reconstruction efficiency required to detect this asymmetry. In Fig.~2
we again set $\delta m^2_{H,A} = 0$. In this case the total rate
asymmetry $A_R(1)$ is entirely due to $h-H$ interference, and is
hence unmeasurably small.  In contrast, near the Higgs peak the
effective polarization asymmetries $\hat{A}_P(1) \propto \{C_6\}$ and
$\hat{A}_P(2) \propto \{C_{12} \}$ are both very large. Recall that
the former can be measured with longitudinally polarized beams, while
the latter can be studied only if at least one beam is transversely
polarized. In Figs.~2 and 3 we have assumed 100\% beam
polarization. Imperfect polarization would dilute these effective
asymmetries linearly. The effective rate asymmetries $\hat{A}_R(5)
\propto [C_{10}]$ and $\hat{A}_R(9) \ \propto [ C_8]$ can also reach
the level of 1 fb$^{1/2}$. Note that the latter goes through zero at
$\rs = m_H$, and falls only slowly away from the pole region. However,
this asymmetry is only measurable with one longitudinally and one
transversely polarized beam; the effective asymmetry therefore scales
like the square of the overall degree of beam polarization, which
means that the luminosity required to see an effect scales as the
inverse fourth power of the degree of polarization. Finally, the
effective polarization asymmetry $\hat{A}_P(3) \propto \{C_{15}\}$
goes through zero at $\rs = m_A$. It drops off less quickly away from
the pole region than the other polarization asymmetries do, but for
this set of parameters it always stays below 0.5 fb$^{1/2}$. All
polarization/angle asymmetries in Figs.~2 and 3 refer to $\ttau_1^+
\ttau_1^-$ production; in some cases the corresponding asymmetries for
$\ttau_1 \ttau_2$ production are even larger.

In Fig.~3 we show results for $\delta m^2_{H,A} = 100$ GeV$^2$
\cite{12}. This increases the mass splitting between the two heavy
Higgs bosons by less than 50 MeV. However, the contributions of these
two bosons can now interfere, since each of them has both scalar and
pseudoscalar couplings to muons. This can lead to a large effective
rate asymmetry $\hat{A}_R(1)$. Note that this asymmetry goes through
zero at a value of \rs\ between the masses of the two Higgs
eigenstates. In contrast, the effective rate asymmetry $\hat{A}_R(9)$
remains almost the same as before; in particular, it still goes
through zero for \rs\ very close to the mass of the heavier (mostly
CP--even) Higgs boson. The measurement of these asymmetries as a
function of the beam energy could therefore allow to determine the
mass splitting between the two heavy Higgs bosons; for the chosen
example studies of the overall line shapes would probably only allow
to give an upper limit on this quantity.

For $\rs \simeq m_A$ the effective polarization asymmetry
$\hat{A}_P(1)$ remains as in Fig.~2. It does change away from the
poles; in particular, its zero moves from $\sim 506$ GeV to $\sim 501.2$
GeV. However, the location of this zero will be difficult to determine,
since $|\hat{A}_P(1)|$ remains small for larger \rs. Finally, $\hat{A}_P(2)$
and $\hat{A}_P(3)$ (not shown) remain essentially the same as for
$\delta m^2_{H,A} = 0$.

In the simple example used here, measurements with unpolarized beams
and with (at least) one longitudinally polarized beam are sufficient
to determine the values of the two fundamental CP--violating
parameters, $\delta m^2_{H,A}$ and the relative phase between $\mu$
and $A_\tau$. Recall, however, that we have assumed that even $\ttau_1
\ttau_2$ can be produced in on--shell Higgs boson decays; if this is
not true, all rate asymmetries will be very small. Also, $\delta
m^2_{H,A}$ could itself be complex \cite{12}, if $\tilde t$ or
$\tilde b$ pairs can be produced in Higgs decays; this would increase
the number of CP--odd parameters by one. Finally, many additional
CP--phases can be introduced if there is nontrivial flavor mixing
among sleptons \cite{11b}. In all these cases the availability of
transversely polarized beams is essential to fully disentangle the
sources of CP--violation. We will discuss these issues in more detail
in a future publication.

\subsection*{Acknowledgements}
The work of SYC was supported in part by the KOSEF--DFG large collaboration
project, Project. No. 96-0702-01-01-2. MD acknowledges financial
support from FAPESP (Brazil).


\clearpage
\noindent

\setcounter{figure}{0}

\vspace*{-2cm}
\begin{figure}[h]
\centerline{\epsfig{file=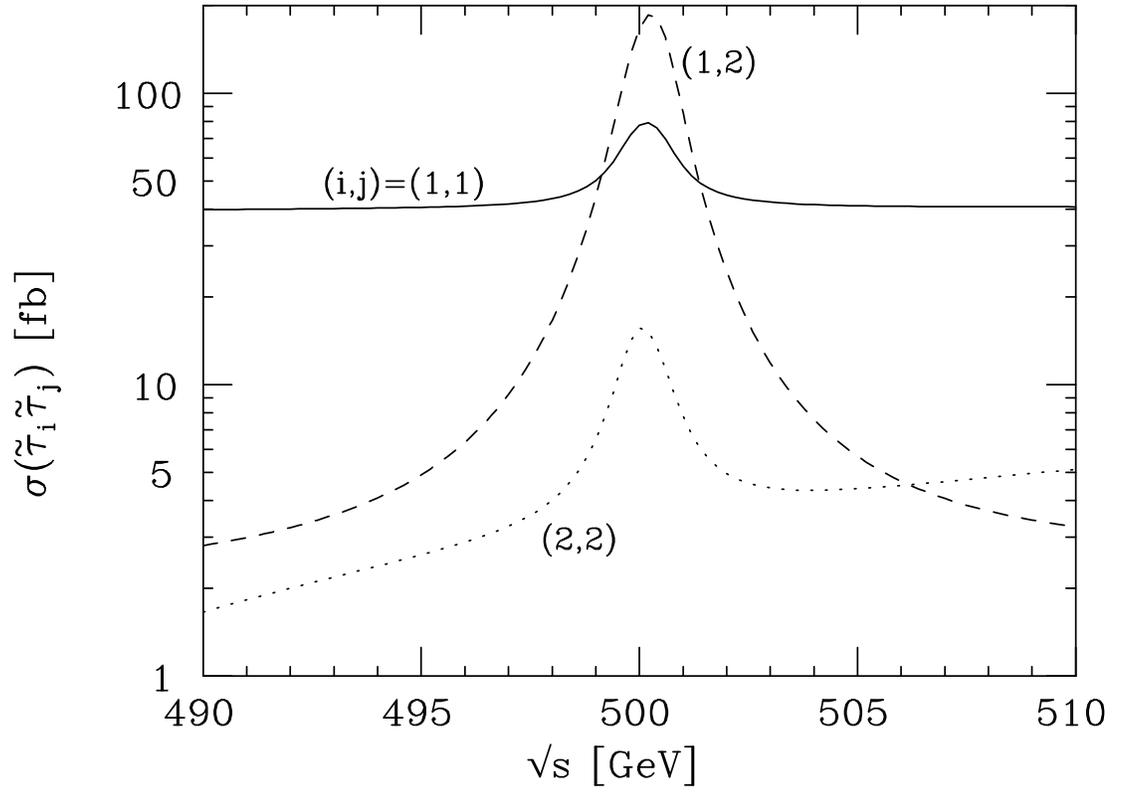,height=17cm}}

\caption
{Total cross sections for $\mumu \rightarrow \ttau_i^- \ttau_j^+$.
The curve labelled ``(1,2)'' refers to the sum of $\ttau_1^- \ttau_2^+$
and $\ttau_1^+ \ttau_2^-$ production. See the main text for the choice
of parameters.}

\end{figure}


\begin{figure}[h]

\vspace*{-7cm}

\centerline{\epsfig{file=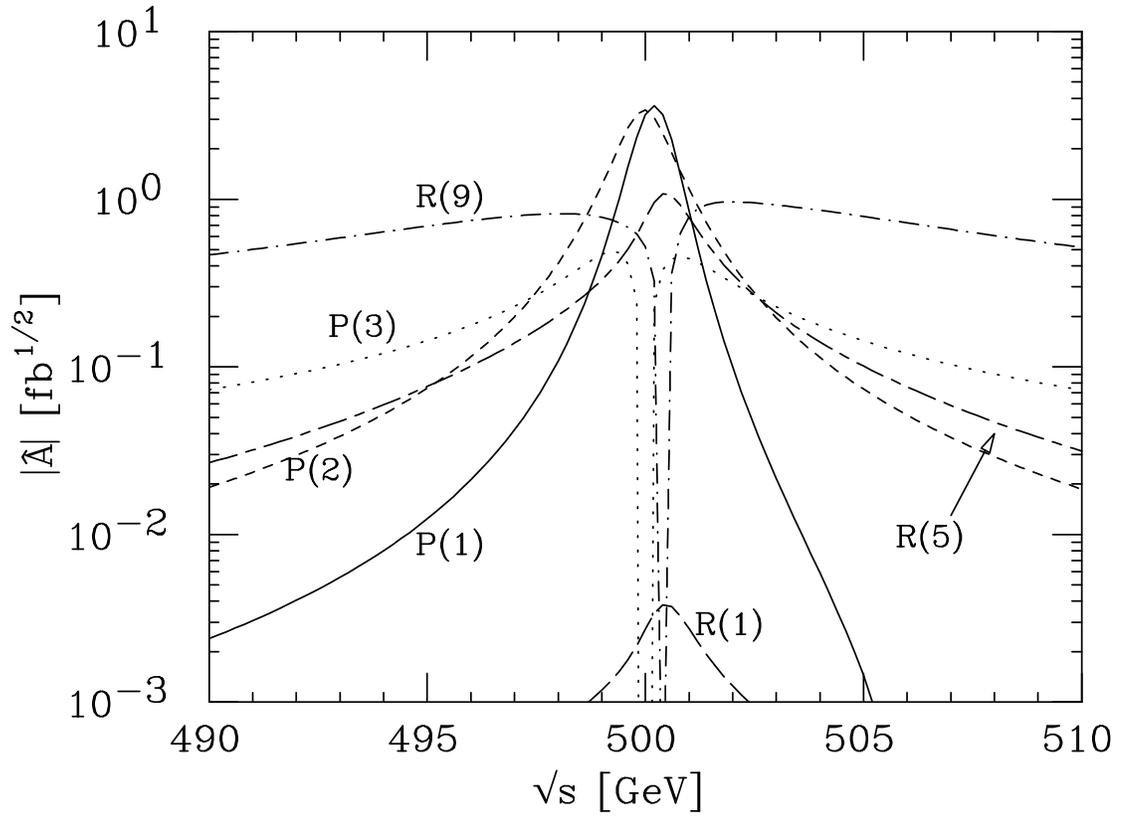,height=17cm}}

\caption
{Absolute values of selected asymmetries times square root of the 
cross section for the same set of parameters as in Fig.~1. The labels
'R' and 'P' refer to rate and polarization/azimuthal angle asymmetries,
respectively.}

\end{figure}

\clearpage
\vspace*{-2cm}

\begin{figure}[h]
\centerline{\epsfig{file=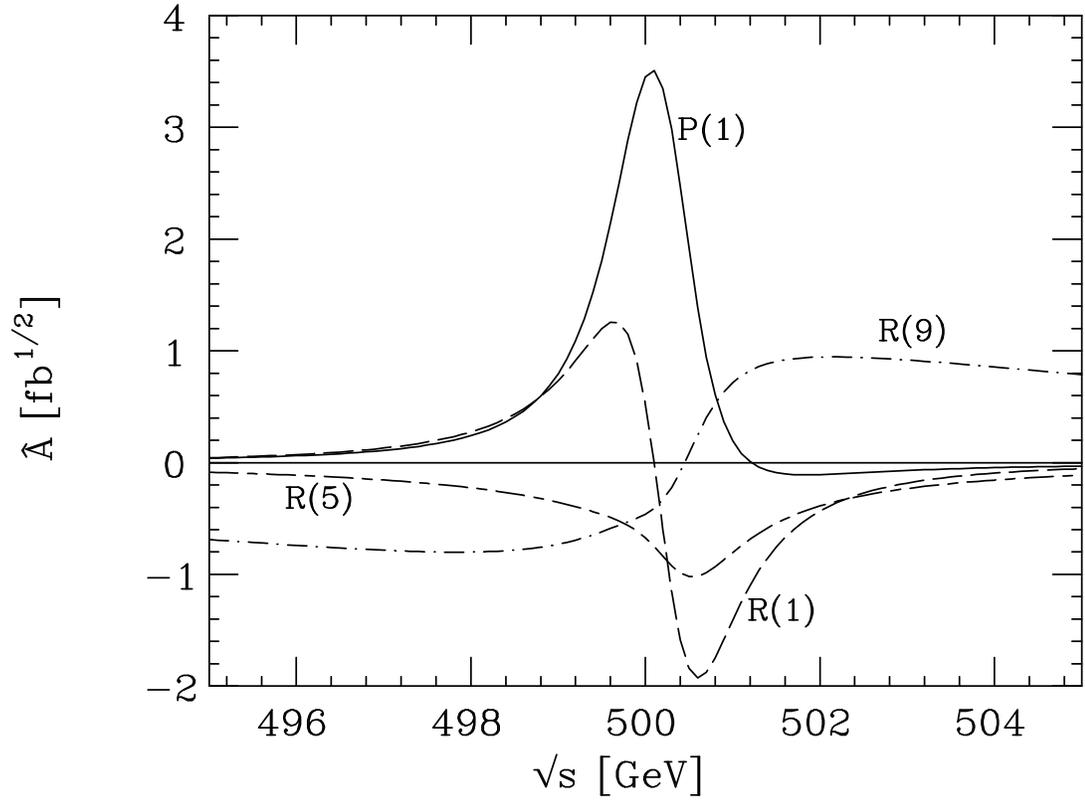,height=17cm}}
\caption
{Selected effective asymmetries for the same parameters as in Fig.~1,
except that we have introduced a nonzero $\delta m_{H,A}^2 = 100$
GeV$^2$.}

\end{figure}

\end{document}